\documentclass[12pt,letterpaper]{article}
\usepackage{epsfig,rotating,setspace,latexsym,amsmath,epsf,amssymb,amsfonts,bm,theorem,cite,caption,subcaption,enumerate,longtable,accents,bbm}
\usepackage{algorithm,algorithmic,graphicx,epsf,authblk,epstopdf,url,color,multirow}
\usepackage{mathtools,comment,tabularx}
\usepackage{comment}
\usepackage[inline]{enumitem}

\usepackage{dsfont}

\newtheorem{theorem}{Theorem}

\newtheorem{corollary}{Corollary}
\newtheorem{definition}{Definition}

\newtheorem{lemma}{Lemma}

\newenvironment{Proof}[1]{\medskip\par\noindent{\bf Proof:\,}\,#1}{{\mbox{\,$\blacksquare$}\par}}

\newcolumntype{Y}{>{\centering\arraybackslash}X}
\newcommand{\figref}[1]{\figurename~\ref{#1}}
\allowdisplaybreaks

\title{ Refined Bitcoin Security-Latency Under Network Delay}
	
\author{Mustafa Doger \qquad Sennur Ulukus\\
\normalsize Department of Electrical and Computer Engineering\\
\normalsize University of Maryland, College Park, MD 20742\\
\normalsize  \emph{doger@umd.edu} \qquad \emph{ulukus@umd.edu}}

\begin{document}
\date{}
\maketitle

\begin{abstract}
     We study security-latency bounds for Nakamoto consensus, i.e., how secure a block is after it becomes $k$-deep in the chain. We improve the state-of-the-art bounds by analyzing the race between adversarial and honest chains in three different phases. We find the probability distribution of the growth of the adversarial chains under  models similar to those in [Guo, Ren; AFT 2022] when a target block becomes $k$-deep in the chain. We analyze certain properties of this race to model each phase with random walks that provide tighter bounds than the existing results. Combining all three phases provides novel upper and lower bounds for blockchains with small $\lambda\Delta$.
\end{abstract}

\section{Introduction}\label{intro-sec}
Introduced by Nakamoto in 2008 \cite{btc-whitepaper}, Bitcoin enables public users to maintain a ledger in a distributive manner. Transactions listed in the public ledger are secured by the longest chain protocol using a Proof of Work (PoW) approach. Honest miners who follow the protocol extend their longest chain of blocks containing the transactions. A new block is required to contain a nonce that satisfies difficulty requirements of the chain, making it valid. However, even if a block is at the tip of the longest chain and satisfies all requirements, it cannot be immediately confirmed due to a phenomenon called ``forking.'' The forking phenomenon occurs due to the network delays and adversarial activities.

Assuming all miners work on the same longest chain, the first one to mine a new block sends the new block to its peers immediately. However, due to the finite speed of light and network constraints, the peers receive the new block with some delay. During this delay, they might be able to mine a new block themselves with different content at the same height, hence the name ``forking.'' Forks also happen when adversaries do not follow the longest chain protocol and try to undermine the ledger. As a result, the deeper a transaction is in the longest chain, the more secure its content will be. This observation hints at the trade-off between security and latency of transactions. Following the Bitcoin whitepaper \cite{btc-whitepaper}, one of the first to study the latency-security problem of blockchains against all possible attacks is Garay et.~al. \cite{garay}. Further studies have extended the analysis to practical non-lockstep protocols, see e.g., \cite{non-lockstep-sec-lat}. 

Recently, several studies have given extensive bounds on the security-latency trade-off under network delay $\Delta$ \cite{guo-close-sec-lat, guo-btc-sec-lat, gazi-sec-lat}. They investigate how secure a block is at any point during the execution of the protocol under a network delay of $\Delta$. Specifically, \cite{guo-close-sec-lat} considers races between Poisson and renewal processes to give upper and lower bounds on how secure a block is after it is confirmed. \cite{gazi-sec-lat} analyzes a dynamic programming algorithm to numerically bound the safety violation probability. While \cite{guo-close-sec-lat} considers confirmation rules that treat latency in terms of time units, \cite{guo-btc-sec-lat} considers the case where the committing rule is defined in terms of how deep a block is in the current longest chain, and outperforms the results of \cite{gazi-sec-lat} in small $\lambda\Delta$ regimes, such as Bitcoin where $\lambda=1/600$ blocks/s and $\Delta\approx 10$ seconds according to the historical data.

Our work focuses on the approach studied in \cite{guo-btc-sec-lat}, and we tighten the lower and upper bounds of security guarantees significantly. We use the ``rigged'' model introduced in \cite{guo-btc-sec-lat} and modify the region of interest where adversarial and honest chains are racing when the target transaction enters the system. By doing so, we present significant improvements, and by orders of magnitude, for certain parameter regimes. We provide a formula to calculate the probability of achievable and converse results. We present our results for Bitcoin and PoW-Ethereum settings with various fractions of adversarial presence.

For Bitcoin, where a block is mined approximately every $10$ minutes and assuming network delay is at max $10$ seconds, under widely adopted $6$-block confirmation rule, safety violation probability is narrowed down to between $0.112\%$ and $0.177\%$ under $10\%$ adversarial presence. For comparison, the previous best-known results \cite{guo-btc-sec-lat} were between $0.106\%$ and $0.353\%$. 

\section{System Model}
\subsection{Protocol: Honest Nodes and Adversaries}
We assume that the reader has familiarity with blockchain protocols. Here, we abstract out the basics of the protocol together with the blockchain data structure and validity constraints. In this abtract system, $n$ nodes participate in a network to maintain a distributed ledger which initially consists solely of the genesis (zeroth) block. Honest nodes, who make up $\alpha$ fraction of all nodes, stick to the protocol, i.e., they try to mine a new block at the tip of the longest chain they have seen so far. Whenever a block is mined by an honest miner, the block is shared and assumed to be seen by all miners within a $\Delta$ amount of time. Adversarial miners are allowed to deviate from the protocol, i.e., they are not required to mine at the tip of their longest chain and can decide not to share their blocks. However, their blocks should contain a valid PoW (or Proof of Stake (PoS) depending on the model). 

A widely adopted model for building a blockchain data structure is to assume that new blocks arrive (i.e., are mined) according to a Poisson process. Hence, the interarrival-times of mined blocks are independent exponentially distributed random variables with mean $1/\lambda$, and a block is honest with probability (w.p.) $\alpha$ by Poisson splitting. The fraction of adversarial miners in the system is denoted by $\beta=1-\alpha$, which is assumed to satisfy $\beta<\frac{1-\beta}{1+(1-\beta)\lambda\Delta}$ \cite{nakamoto-always-wins}. We further assume that the entire adversarial power is concentrated in the hands of a single entity and adopt the convention of strong adversaries in the literature, i.e., the adversary controls the network delay as long as any introduced delay is at most $\Delta$ and the ties are broken in the adversary's favor. 

\subsection{Confirmation Rule}
A block and transactions within that block are considered to be confirmed according to a $k$-block confirmation rule, if it is part of the longest chain and there are at least $k-1$ blocks mined on top of it, in the view of an honest miner. The aim of the adversary is to spend the same resource on more than one transaction, i.e., double spend. Hence, we say a confirmed transaction is discarded if and only if a block containing the transaction is confirmed, and later, another block containing a conflicting transaction on the same height is confirmed.

In this paper, we are interested in a certain ``target transaction'' $tx$ which enters the transaction pool at time $\tau$. We assume that honest miners try to mine a new block (``target block'') containing $tx$ at the tip of their longest chain if possible. We would like to calculate lower and upper bounds on the probability of discarding $tx$ after confirmation. In this terminology, anything that the adversary has the ability to do is considered achievable (lower bound). For the upper bound, as explained in \cite{guo-btc-sec-lat}, we use a ``rigged'' model which makes the adversary strictly more powerful than physically possible, hence an unachievable scenario.

\subsection{Parameters and Definitions}
In this sub-section, we provide some frequently-used parameters and definitions.

\begin{definition} {\normalfont \textbf{(Jumper)}} Jumpers are the first honest blocks that are mined at least $\Delta$ time after each other starting with the genesis block, which is the zeroth jumper \cite{guo-close-sec-lat}.
\end{definition}

\begin{definition}
    {\normalfont \textbf{(Rigged block)}} A block mined by an honest (adversarial) node is simply called an honest (adversarial) block. In this sense, a rigged block is an honest block that is treated as if it is an adversarial block \cite{guo-btc-sec-lat}. The adversary can use the rigged block in any way it wants in order to increase the chance of double spend.
\end{definition}

\begin{definition}
    {\normalfont \textbf{(Block publication)}} We say a block becomes public when it enters the view of every honest miner. Any honest block mined at $t$, becomes public by $t+\Delta$, i.e., enters the view of all honest nodes by $t+\Delta$.
\end{definition}

\begin{table}[h!]    
    \begin{center}
\begin{tabular}{||c c||} 
 \hline
 Parameters & Values \\ [0.5ex] 
 \hline\hline
 $\alpha_{i}$ & $\alpha\cdot e^{-\beta\lambda\Delta}\cdot \frac{(\beta\lambda\Delta)^{i}}{i!}$ \\ 
 \hline
 $\beta_{1}$ & $1-\alpha_{0}-\alpha_{1}$ \\
 \hline
 $\bar{\alpha}_{i}$ & $\alpha\cdot e^{-\lambda\Delta}\cdot \frac{(\lambda\Delta)^{i}}{i!}$ \\
 \hline
 $\bar{\alpha}$ & $\alpha e^{-\lambda\Delta}$  \\
 \hline
 $\rho$ & $\alpha e^{-\lambda\Delta}\cdot(1+\lambda\Delta+\beta-\alpha e^{-\lambda\Delta})$  \\
 \hline
 &  \\   [-1.1em]
 $\bar{\beta}^2$ & $1-\bar{\alpha}^2-\rho$\\ 
 \hline
  $\Bar{a}_i$ & $\sum_{j\geq i}\Bar{\alpha}_j + \beta\cdot\mathbbm{1}_{i\leq 2}$\\[0.5ex]
 \hline
  & \\[-2ex]
  $\Bar{b}_i$ & $\sum_{j\geq i}\Bar{\alpha}_j + \beta\cdot\mathbbm{1}_{i\leq 1}$\\[0.5ex]
 \hline
  $a_i$ & $\sum_{j\geq i}\alpha_j + \beta\cdot\mathbbm{1}_{i\leq 2}$\\[0.5ex]
 \hline
  $b_i$ & $\sum_{j\geq i}\alpha_j + \beta\cdot\mathbbm{1}_{i\leq 1}$\\[0.5ex]
 \hline
\end{tabular}
\end{center}
\caption{ Frequently used notations.}
\label{freq-table}
\end{table}

\section{Lower Bound}\label{lower-section}
There are different achievable adversarial strategies. The strategy considered in \cite{guo-btc-sec-lat} makes use of $\Delta=0$ which is a lower bound to non-zero delay since the adversary can always choose not to introduce any delays. Depending on the magnitude of $\lambda\Delta$, however, this bound may actually be quite off from reality. Here, we deploy an adversarial strategy known as \emph{private attack} with $\Delta$ delay, a simple yet effective one to hamper the growth of the honest chain. Specifically, adversary delays every honest block by the maximum allowed $\Delta$, in the meanwhile, mining a private chain in order to double spend. As the adversary controls the network delay, in this context, a block published $\Delta$-delay after its mining time, $\tau_m$, means that the block becomes public exactly at $\tau_m+\Delta$.

In this scenario, the length of the honest chain will be equal to the number of published jumpers from the start of the protocol until the current time.  Hence, under the $k$-block confirmation rule, we calculate a lower bound for the probability that the  $(k-1)$st jumper after the target block is mined and at some point from then on $tx$ is excluded from the longest chain of an honest miner. Our analysis extends the one in \cite{guo-btc-sec-lat} by incorporating the effects of $\Delta$ delay. 

\subsection{Pre-Mining Gain}
We start with the pre-mining gain, i.e., the lead $L$, which is the difference in the heights of the longest chain versus the longest honest chain before $\tau$; this is the adversary's lead. Assuming $\tau$ is large enough, the lead can be modeled as the steady state distribution of an extended birth-death process \cite{guo-btc-sec-lat, guo-close-sec-lat}. When the birth-death process is extended to incorporate the $\Delta$ delay strategy, we obtain a Markov chain with the following state transition probability matrix,
\begin{align}
    P=\begin{bmatrix}
        \alpha_{0} & \alpha_{1}+\beta & \alpha_{2}  & \alpha_{3} & \alpha_{4} & \ldots\\
        \alpha_{0} & \alpha_{1}       & \alpha_{2}+\beta & \alpha_{3} & \alpha_{4} & \ldots\\
        0          & \alpha_{0} & \alpha_{1} & \alpha_{2}+\beta & \alpha_{3} & \ldots\\
        0          & 0 & \alpha_{0} & \alpha_{1} & \alpha_{2}+\beta  &  \ldots\\
        \vdots & \vdots & \vdots & \vdots & \vdots & \ddots \\
    \end{bmatrix} \label{pmatrix}
\end{align}

 This Markov chain is constructed as follows: At each ``mining event,'' we toss a coin, the result is either an adversarial block w.p.~$\beta$ or an honest block w.p.~$\alpha$, in which case we further consider how many adversarial arrivals there have been in the $\Delta$ delay interval of that honest block, and ignore the honest arrivals during that $\Delta$ delay. Hence, we assume that any honest block arrives together with some number of adversarial blocks which are mined within the $\Delta$ delay of the honest block, and that number has a Poisson distribution. In (\ref{pmatrix}), the $(i+1,j+1)$th element of $P$ denotes the probability that the lead goes from $i$ to $j$. Here, $\alpha_{i}$ is the probability that an honest block is mined (which happens w.p.~$\alpha$) followed by $i$ adversarial blocks are mined within the $\Delta$ delay interval (which happens w.p.~$e^{-\beta\lambda\Delta}\cdot \frac{(\beta\lambda\Delta)^{i}}{i!}$). It is worth noting that, in this process, when one considers only honest blocks mined $\Delta$-time after each other, and ignores honest blocks (or if they get rigged) that are mined within $\Delta$-delay, all honest blocks effectively become jumper blocks. Thus, in this paper we use honest and jumper blocks interchangeably. Note that, the key point in this process is to identify renewal times, where each arrival is either an adversarial block, or a jumper together with the $\Delta$ delay that ensues it.

\begin{figure}[t]
	\centerline{\includegraphics[width=0.8\columnwidth]{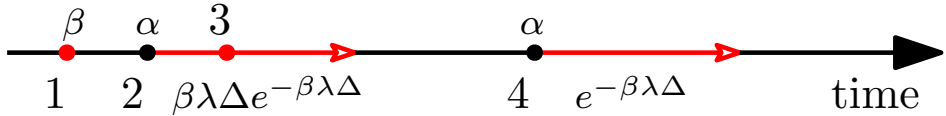}}
	\caption{A sample path of arrivals.}
	\label{lower-seqn}
\end{figure}

As an explicit example, consider the $(1,2)$th element of $P$, which represents the probability that the lead goes from $0$ to $1$. This happens in two different ways:  First, w.p.~$\alpha_1$ one honest block is mined and one adversarial block is mined within $\Delta$ delay of that honest block. Since adversary is aware of all honest blocks with no delay, it will build its block on top of the new honest block, i.e., the lead increases from $0$ to $1$. Second, w.p.~$\beta$ the mined block is adversarial and no honest block is mined. Again, this is a net $1$ gain for the adversary. Consequently, the $(1,2)$th element of $P$ is $\alpha_1+\beta$. In addition, \figref{lower-seqn} shows a sample path of arrivals to visualize this random process. The first coin toss results in an adversarial block $1$ w.p.~$\beta$. The second coin toss, independent of the first one, results in an honest block $2$ w.p.~$\alpha$, and we consider the $\Delta$ delay interval of this toss (represented by the red arrow) which results in an additional adversarial block w.p.~$\beta\lambda\Delta e^{-\beta\lambda\Delta}$. Any honest arrivals during this red interval can be ignored, as they only fork the jumper chain, and do not contribute to the length of the longest chain. After the $\Delta$ delay ends, the next coin toss is independent of anything that happened before, hence, we have a memoryless process, represented by a Markov chain. 

 The steady state of matrix $P$ can be found by Ramaswami's formula \cite{ramaswami-mg1}, which we provide at Appendix~\ref{app-a}. Since we are finding a lower bound for the lead, we further simplify this strategy by assuming that there are at most two adversarial arrivals during the $\Delta$ delay interval of any jumper, which essentially truncates the Poisson distribution during the $\Delta$ delay interval. Thus, there are three possible outcomes of a coin toss that represents the next arrival: 
\begin {enumerate*} 
\item $E_{1}$, denoting a jumper honest block arrival (no adversarial arrival during the $\Delta$ delay interval) results in honest chain growing by one block.
\item $E_{2}$, denoting a jumper honest block arrival together with an adversarial block arrival during the $\Delta$ delay interval that results in one block of growth in both honest and private chains.
\item $E_{3}$, denoting an adversarial arrival or a jumper honest block arrival together with two adversarial block arrivals during the $\Delta$ delay interval.
\end{enumerate*} 
Note that the two possible events listed in $E_{3}$ result in net one block growth in the adversarial chain, so we treat them the same. Furthermore, if the lead is zero, for the sake of simplicity, we assume that there can be at most one adversarial arrival during the $\Delta$ delay interval of a jumper. This simplification means that we merge $E_{3}$ and $E_{2}$ when the lead is zero, which is a valid lower bound. Given the above arguments, the following Markov chain is a lower bound on the lead of the best adversarial strategy,
\begin{align}
    P'=\begin{bmatrix}
        \alpha_{0} & 1-\alpha_{0} & 0  & 0 & 0 & \ldots\\
        \alpha_{0} & \alpha_{1}   & \beta_{1}  & 0 & 0 & \ldots\\
        0          & \alpha_{0}   & \alpha_{1} & \beta_{1} & 0 & \ldots\\
        0          & 0 & \alpha_{0} & \alpha_{1} & \beta_{1} & \ldots\\
        \vdots & \vdots & \vdots & \vdots & \vdots & \ddots \\
    \end{bmatrix}
\end{align}
 Letting $\pi_i$ denote $\mathbb{P}(L=i)$, we find  $\pi_{0}=\frac{\alpha_{0}-\beta_{1}}{1-\beta_{1}}$, $\pi_{1}=\pi_{0}\cdot\frac{1-\alpha_{0}}{\alpha_{0}}$, and $\pi_{i}=\pi_{1}(\frac{\beta_{1}}{\alpha_{0}})^{i-1}$ for $i>1$. We denote the PMF of $L$ with $P_1$.

\subsection{Confirmation Interval}
At time $\tau$, which is assumed to be large and random, target transaction $tx$ enters the transaction pool of miners and will be included in the next honest block by assumption. However, $\tau$ may be in the $\Delta$ delay immediately following a jumper that arrived before $\tau$. In this case, since we assume that the adversary applies the $\Delta$-delay strategy, if an honest block containing $tx$ is mined during this $\Delta$-delay, it forks the last jumper of pre-mining phase and becomes void. Thus, $tx$ is effectively included in the honest chain after the end of the $\Delta$ delay immediately following the last jumper of the pre-mining phase. Hence, without loss of generality, we can assume that $tx$ enters the mempool at a time $\tau$, where the first honest block to be mined contains the $tx$ and is a jumper. Further, this observation implies that the steady-state distribution of $P'$ is a lower bound for the adversarial lead. We call the interval starting at $\tau$  and ending when the target block containing $tx$ becomes $k$-deep in the longest chain of all honest miners, the confirmation interval. Our goal is to find the number of adversarial blocks mined during this confirmation interval. 

First, note that the lead found previously is a steady-state scenario and assumed to be independent of the Poisson arrivals starting at $\tau$,  i.e., $\tau$ is large. Second, statistically, it does not matter if the confirmation interval ends when $tx$ becomes $k$-deep in any honest view or all honest views under the $\Delta$ delay strategy. To see this second point, let $\tau_c$ denote the mining time of the $k$th jumper block mined starting from $\tau$. Then, the former interval (i.e., the interval when $tx$ becomes $k$-deep in any honest view) refers to $[\tau,\tau_c]$ whereas the latter interval (i.e., the interval when $tx$ becomes $k$-deep in all honest views) refers to  $[\tau,\tau_c+\Delta]$. If the adversarial power can discard the target $tx$ in the former interval, it can do so in the latter interval as well, since the latter interval contains the former interval. If discarding happens in the latter interval, but not in the former interval, then, this necessarily implies that the adversary has mined some blocks during the $\Delta$ delay of the $k$th jumper block to catch the longest honest chain. However, if we were to use the former definition, we would have to consider the $\Delta$ delay interval of the $k$th jumper block separately right after the confirmation interval, since only the adversarial chain can grow during the $\Delta$ delay of the $k$th jumper block, ending up with the same result only to make the computations unnecessarily complex. Hence, we use the interval $[\tau,\tau_c+\Delta]$ as our confirmation interval.

As a side note, a similar interval is considered in \cite{guo-btc-sec-lat}, i.e., a binomial random variable with $2k-L$ tosses, which could also be done with a Pascal random variable representing the number of adversarial blocks mined during the confirmation interval instead. This observation leads us to find a similar arrival distribution for the confirmation interval defined above. 

Let $C$ denote the number of adversarial blocks mined between the publication times of two consecutive jumper blocks. In Lemma~\ref{lemma-lower-geo} below, we find the PMF of $C$ by first considering the number of adversarial arrivals before the arrival of the honest block and add it to the number of adversarial arrivals during the $\Delta$ delay interval.

\begin{lemma}\label{lemma-lower-geo}
Under the $\Delta$ delay strategy, the number of adversarial blocks mined between the publication times of two consecutive jumper blocks $C$ has the following distribution,
\begin{align}
    P_{C}(c)=\alpha\beta^{c} e^{-\beta\lambda\Delta}\sum_{j=0}^{c}\frac{(\lambda\Delta)^j}{j!}
    \label{lower_geo_eq}
\end{align}
\end{lemma}

\begin{Proof}
After an honest (jumper) is published, we know that the probability of having $m$ adversarial arrivals before the next honest arrival (which is a jumper by definition) is $\beta^m\alpha$. However, after this honest block arrives, there is a $\Delta$ delay interval until it is published, during which all honest arrivals will be ignored, whereas there can be additional $j$ adversarial arrivals w.p.~$e^{-\beta\lambda\Delta}\frac{(\beta\lambda\Delta)^j}{j!}$. Combining these two results, i.e., $c=j+m$, we obtain \eqref{lower_geo_eq}. Clearly, at the end of the $\Delta$ delay interval, this jumper will be published, and thus, the next jumper also has the same distribution and they are independent by the memoryless property, i.e., they are i.i.d.
\end{Proof}

Let $S$ denote the number of adversarial arrivals during the confirmation interval. Next, Corollary~\ref{corollary-lower-pascal} finds the PMF of $S$ using the PMF of $C$ found in Lemma~\ref{lemma-lower-geo}. To do so, we first express $C$ in terms of its probability generating function (PGF), note that PGF of $S$ is the $k$th power of the PGF of $C$, and finally transform the result back to the time domain.

\begin{corollary}\label{corollary-lower-pascal}
The number of adversarial arrivals during the confirmation interval $S$ has the following distribution,
\begin{align}
    P_{S}(s) &= \alpha_{0}^{k}\beta^s\sum_{n=0}^{s}\binom{k\!-\!1\!+\!n}{n}\frac{(\lambda\Delta k)^{s-n}}{(s-n)!} \label{lower-pascal-like-eqn}
\end{align}
\end{corollary}

\begin{Proof}
All jumpers including the first has the distribution of $C$ in (\ref{lower_geo_eq}). Further, $S$ is the sum of $k$ i.i.d.~random variables each distributed as $C$. First, we find the PGF of $C$, i.e., $G_C(z)=\mathbb{E}[z^C]$, as
\begin{align}
    G_{C}(z)&=\alpha e^{-\beta\lambda\Delta}\sum_{i\geq0}(z\beta)^{i}\sum_{j=0}^{i}\frac{(\lambda\Delta)^j}{j!}\\
    &=\alpha_{0}\sum_{j\geq0}\sum_{i\geq j}(z\beta)^{i}\frac{(\lambda\Delta)^j}{j!}\\
    &=\alpha_{0}\sum_{j\geq0}\frac{(z\beta\lambda\Delta)^j}{(1-z\beta)\cdot j!}\\
    &=\alpha_{0}\frac{e^{z\beta\lambda\Delta}}{1-z\beta}
\end{align}
Next, we find the PGF of $S$, as
\begin{align}
    G_{S}(z) &= G_{C}(z)^{k}\\
    &=\alpha_0^{k}\left(\frac{e^{z\beta\lambda\Delta}}{1-z\beta}\right)^{k} \label{intermed1}\\
    &=\alpha_0^{k}\sum_{n\geq0}\binom{k\!-\!1\!+\!n}{n}(z\beta)^{n}\sum_{m\geq0}\frac{(z\beta\lambda\Delta k)^m}{m!} \label{intermed2}\\
    &=\alpha_0^{k}\sum_{m\geq0}\sum_{n\geq0}\binom{k\!-\!1\!+\!n}{n}\frac{(\lambda\Delta k)^m}{m!}(z\beta)^{n+m}\\
    &=\sum_{s\geq0}\alpha_0^{k}(z\beta)^{s}\sum_{n=0}^{s}\binom{k\!-\!1\!+\!n}{n}\frac{(\lambda\Delta k)^{s-n}}{(s-n)!} \label{intermed3}
\end{align}
where in going from (\ref{intermed1}) to (\ref{intermed2}), we use the identity,
\begin{align}
    \sum_{n\geq0}\binom{k\!-\!1\!+\!n}{n}x^{n}=\frac{1}{(1-x)^{k}}
    \label{pascal}
\end{align}
The desired result follows from (\ref{intermed3}). 
\end{Proof}

We denote the PMF of $S$ with $P_2$. At this point, it should be clear that the adversary wins the race if sum of the lead and confirmation gain is more than $k$, i.e., $L+S \geq k$. If not, the race enters the last phase called the post-confirmation race.

\subsection{Post-Confirmation Race}
After the confirmation interval, if the adversary is behind in the race, it still has a chance to win. Let $D=k-L-S$ denote the deficit of the adversary right after the confirmation interval. We can represent this part of the race with a random walk that starts from the origin and moves according to a three-way coin toss with events $E_1$ (moving one step to the left), $E_2$ (net zero movement), and $E_3$ (moving one step to the right). We can denote the $i$th toss with $W_i\in\{-1,0,1\}$. Note that we trim the adversarial arrivals during the $\Delta$ delay interval to at most two as we did in the pre-mining gain stage. Let $T_{i}$ denote the current position of this random walk after $i$ three-way coin tosses, i.e., $T_i= \sum_{j=1}^{i}W_j$. If the random walk ever reaches $D$, i.e., $\max\limits_{i\geq1} T_i \geq D$, then, the adversary wins. Moreover, if $\max\limits_{i\geq1} T_i = D-1$ and $E_2$ happens while the random walk is at $D-1$ at any point of the process, the adversary wins due to the ability of publishing adversarial blocks mined during the $\Delta$ delay interval before the jumper. Thus, combining these two possibilities, we denote the event that the adversary catches the honest chain with $\max\limits_{i\geq1} T'_i=\max\limits_{i\geq1} \left(T_i+\mathds{1}_{W_i= 0} \right)\geq D$. In Lemma~\ref{deficit-lower-lemma} we find the probability of this event. First, we prove the following intermediate results in Lemma~\ref{random-walk-num-of-max-lemma} and Corollary~\ref{corollary2}.

\begin{lemma}\label{random-walk-num-of-max-lemma}
Consider a binary random walk starting from the origin and moving to the left w.p.~$\alpha$ and right w.p.~$\beta<\frac{1}{2}$. Let $M$ denote the maximum point the random walk ever reaches. Let $N$ denote the number of times the random walk reaches the point $M$. Then,
\begin{align}
    \mathbb{P}(N=n|M=m)=\alpha\cdot\beta^{n-1}, \quad n\geq 1
\end{align}
\end{lemma}

\begin{Proof}
 We start by analyzing the case $n=1$, and observe that for $N=n+1$, the result is recursive by the Markov property. Let $B_{i}$ denote the position after the $i$th coin toss. Let $I$ denote the event that $\max\limits_{i\geq1} B_{i}\geq m$. Note that $\max\limits_{i\geq1} B_{i}$ has geometric distribution \cite[eqn.~(7.3.5)]{ross-stochastic}. Further, given the event $I$, let $s$ denote the time of the first arrival on $m$ and $J_s$ denote the event that $\max\limits_{i>s} B_{i}< m$. Then, $\mathbb{P}(N=1|M=m)$ can be expressed as
\begin{align}
 \mathbb{P}(N=1|M=m) &=\frac{\mathbb{P}(I)\mathbb{P}(B_{s+1}=m-1|I)\mathbb{P}(J_s|B_{s+1}=m-1)}{\mathbb{P}(M=m)} \\
    &=\frac{\left(\frac{\beta}{\alpha}\right)^m\cdot \alpha \cdot \left(1-\frac{\beta}{\alpha}\right)}{\left(1-\frac{\beta}{\alpha}\right)\left(\frac{\beta}{\alpha}\right)^m} \\
    &=\alpha \label{cond-max-num-hit-eq}
\end{align}
For the case $n=2$, we first observe that $\mathbb{P}(N=2|M=m)=\mathbb{P}(N\neq1|M=m)\mathbb{P}(N=2|N\neq1,M=m)$. Also note, that \eqref{cond-max-num-hit-eq} does not depend on $m$. Hence, after $s$, the random walk returns to $m-1$, and from that point reaching $m$ again only once, has a similar calculation, i.e., $\mathbb{P}(N=2|N\neq1,M=m)=\mathbb{P}(N=1|M=m)=\alpha$. Thus, $\mathbb{P}(N=2|M=m)=\mathbb{P}(N\neq1|M=m)\mathbb{P}(N=2|N\neq1,M=m)=\beta\alpha$. The rest of the cases are similar.
\end{Proof}

\begin{corollary}\label{corollary2}
Consider a random walk starting from the origin with a three-way coin toss, each toss denoted by $W_i$, $i\geq1$, and $\mathbb{P}(W_i=-1)=\mathbb{P}(E_1)=\alpha_0$, $\mathbb{P}(W_i=0)=\mathbb{P}(E_2)=\alpha_1$, $\mathbb{P}(W_i=1)=\mathbb{P}(E_3)=\beta_1$,  where $\beta_1<\alpha_0$. Let $M$ denote the maximum point the random walk ever reaches. Let $N$ denote the number of times the random walk moves from $M-1$ to $M$. Then,
\begin{align}
    \mathbb{P}(N=n|M=m)=\frac{\alpha_0}{\alpha_0+\beta_1}\cdot\left(\frac{\beta_1}{\alpha_0+\beta_1}\right)^{n-1}
\end{align}
\end{corollary}

\begin{lemma}\label{deficit-lower-lemma}
Consider a sequence of i.i.d.~random variables denoted by $W_i$, $i\geq1$, and $\mathbb{P}(W_i=-1)=\mathbb{P}(E_1)=\alpha_0$, $\mathbb{P}(W_i=0)=\mathbb{P}(E_2)=\alpha_1$, $\mathbb{P}(W_i=1)=\mathbb{P}(E_3)=\beta_1$ where $\beta_1<\alpha_0$.  Let $T'_i= \mathds{1}_{W_i= 0}+\sum_{j=1}^{i}W_j$. Then, for $a\geq1$,
\begin{align}
    \mathbb{P}\left(\max\limits_{i\geq1} T'_i\geq a\right)=\left(\frac{\beta_1}{\alpha_0}\right)^{a-1}\left(\frac{1-\alpha_0}{1-\beta_1}\right)\label{equation-random-walk-max}
\end{align}
\end{lemma}

\begin{Proof}
We first find the distribution of $\max\limits_{i\geq1} T_i$, which is geometric. Next, to consider the event of $E_2$ while $\max\limits_{i\geq1} T_i = D-1$, we use the result of Corollary~\ref{corollary2}, which gives the distribution of the number of times $\max\limits_{i\geq1} T_i = D-1$. Finally, we consider whether $E_2$ happened or not each time the walk hits $\max\limits_{i\geq1} T_i = D-1$, and combine the results to complete the proof.

Since by definition $T_i \leq T'_i \leq T_i + 1$, we have
\begin{align}
     \max\limits_{i\geq1} T_i\leq\max\limits_{i\geq1} T'_i\leq \max\limits_{i\geq1} T_i +1 \label{equation-bounds-on-max}
\end{align}
Let $Q_a$ denote the following event
\begin{align}
    \max\limits_{i\geq1} T_i\geq a \enspace \cap \enspace \max\limits_{i\geq1} T'_i\geq a=\max\limits_{i\geq1} T_i\geq a
\end{align}
and let $R_a$ denote the following event
\begin{align}
    \max\limits_{i\geq1} T_i=a-1 \enspace \cap \enspace \max\limits_{i\geq1} T'_i=a
\end{align}
Thus, $\mathbb{P}(Q_a)+ \mathbb{P}(R_a)=\mathbb{P}\left(\max\limits_{i\geq1} T'_i\geq a\right)$.

 Let $\mathbb{P}(0\rightarrow1)=p$, where $0\rightarrow1$ denotes the event of reaching state $1$ starting from the origin in a three-way coin toss model. Using the Chapman-Kolmogorov equations,
\begin{align}
    p=\alpha_0 p^2 + \alpha_1 p + \beta_1
\end{align}
Hence, we have
$\mathbb{P}\left(\max\limits_{i\geq1} T_i\geq 1\right)=p=\frac{\beta_1}{\alpha_0}$.
Applying this recursively, we obtain
\begin{align}
\mathbb{P}(Q_a)=\mathbb{P}\left(\max\limits_{i\geq1} T_i\geq a\right)=p^a, \quad a\geq 1
\label{q_a_three rand walk-intermed}
\end{align}
From (\ref{q_a_three rand walk-intermed}), we also conclude, using $\mathbb{P}(Q_a)-\mathbb{P}(Q_{a+1})$, that
\begin{align}
    \mathbb{P}\left(\max\limits_{i\geq1} T_i = a\right)=
    \left(1-\frac{\beta_1}{\alpha_0}\right)\left(\frac{\beta_1}{\alpha_0}\right)^{a} \label{q_a_three rand walk}
\end{align}
We next find $\mathbb{P}(R_a)$. Let $s$ denote a time when the three-way random walk is at the maximum point $a-1$. Then,
\begin{align}
    \mathbb{P}\left(T_{s+1}= a\!-\!2\Big|T_{s}=a\!-\!1, \max\limits_{i\geq s} T_i=a\!-\!1,\max\limits_{i< s} T_i\leq a\!-\!1\right)
     &=\frac{\alpha_0}{\sum\alpha_1^i\alpha_0}\\ &=1-\alpha_1\label{allposib-tprime}
\end{align} 
To understand \eqref{allposib-tprime}, note that the random walk after time $s$ has to return to $a-2$ eventually, since the maximum is $a-1$. However, before going back some tosses might result in $W_i=0$, hence, we consider all possibilities in the denominator.

If the random walk returns to $a-2$ immediately after every time it reaches $a-1$, then, $\max\limits_{i\geq1} T_i$ will be the same as $\max\limits_{i\geq1} T'_i$, since $E_2$ will never happen at the maximum point $a-1$. Thus,
\begin{align}
    \mathbb{P}\left(\max\limits_{i\geq1} T'_i\neq a|\max\limits_{i\geq1} T_i=a-1\right)
    &= \sum_{n=1}^{\infty}\mathbb{P}\left(N=n|\max\limits_{i\geq1} T_i=a-1\right) \cdot(1-\alpha_1)^n\\
    &= \sum_{n=1}^{\infty}\frac{\alpha_0}{\alpha_0+\beta_1}\cdot\left(\frac{\beta_1}{\alpha_0+\beta_1}\right)^{n-1} \cdot(1-\alpha_1)^n \label{intermed4}\\
    &=\frac{\alpha_0}{1-\beta_1} \label{intermed5}
\end{align}
where (\ref{intermed4}) follows from Corollary~\ref{corollary2}, and (\ref{intermed5}) follows from $\alpha_0+\alpha_1+\beta_1=1$. Combining (\ref{intermed5}) with \eqref{q_a_three rand walk} gives
\begin{align}
     \mathbb{P}(R_a)&=\left(1-\frac{\beta_1}{\alpha_0}\right)\left(\frac{\beta_1}{\alpha_0}\right)^{a-1}\frac{\alpha_1}{\alpha_0+\alpha_1} \label{intermed6}
\end{align}
Then, summing $\mathbb{P}(Q_a)$ in (\ref{q_a_three rand walk-intermed}) and $\mathbb{P}(R_a)$ in (\ref{intermed6}) gives the desired result in (\ref{equation-random-walk-max}).
\end{Proof}

As mentioned previously, $\max\limits_{i\geq1} T'_i \geq k-L-S$ refers to the event that the adversary catches the honest chain during the post-confirmation race. We note that PMF of $\max\limits_{i\geq1} T'_i$ and $L$ coincide, hence we denote it with $P_1$ as well. Finally, we put everything together, namely, the analysis of the steady state (i.e., pre-mining gain), confirmation interval, and post-confirmation race in the following theorem.

\begin{theorem}
Given mining rate $\lambda$, honest fraction $\alpha$, delay bound $\Delta$ and confirmation depth $k$, a confirmed transaction can be discarded w.p.~at least:
\begin{align}
    1-\sum_{i+j+l<k} P_1(i)P_2(j)P_1(l) \label{thm-1 eq}
\end{align}
where 
\begin{align*}
    P_1(0)=\frac{\alpha_{0}-\beta_{1}}{1-\beta_{1}},
\end{align*}     
\begin{align*}
    P_1(i)=P_1(0)\cdot\frac{1-\alpha_{0}}{\alpha_{0}}\cdot(\frac{\beta_{1}}{\alpha_{0}})^{i-1},\quad i\geq 1,
\end{align*}
\begin{align*}
    P_{2}(j)= \alpha_{0}^{k}\beta^j\sum_{n=0}^{j}\binom{k\!-\!1\!+\!n}{n}\frac{(\lambda\Delta k)^{j-n}}{(j-n)!},\quad j\geq 0
\end{align*}
 where $\beta_1<\alpha_0$.
\end{theorem}

\section{Upper Bound}
To find an upper bound on the security-latency guarantee, one has to know the best adversarial strategy.  It is proven in \cite{nakamoto-always-wins} that there is no deterministic adversarial strategy that outperforms other strategies for all possible arrivals under non-zero network delay; specifically, see \cite[Lemma F.~1]{nakamoto-always-wins}. However, it is also proven in \cite{guo-btc-sec-lat} that \emph{private attack} is the best attack under the condition that all honest blocks are on different heights (we denote this condition in short as AHBODH). To make use of this fact, \cite{guo-btc-sec-lat} uses a rigged model where some honest arrivals are converted into adversarial ones. The bound performs well for small $\lambda\Delta$ and $k$, however, as these parameters grow, the gap between the upper and lower bounds grows significantly. We refine this model by decreasing the number of rigged blocks without violating the AHBODH condition. We first modify $P$ matrix to incorporate the effect of the rigged blocks for the calculation of the lead. Next, we modify Lemma~\ref{lemma-lower-geo} and Corollary~\ref{corollary-lower-pascal} to incorporate the effect of the rigged model during the confirmation interval. Finally, for the post-confirmation race, we consider two arrivals at a time which brings improvement to the model of \cite{guo-btc-sec-lat}.

\subsection{Pre-Mining Gain}
\begin{figure}[t]
	\centerline{\includegraphics[width=1\columnwidth]{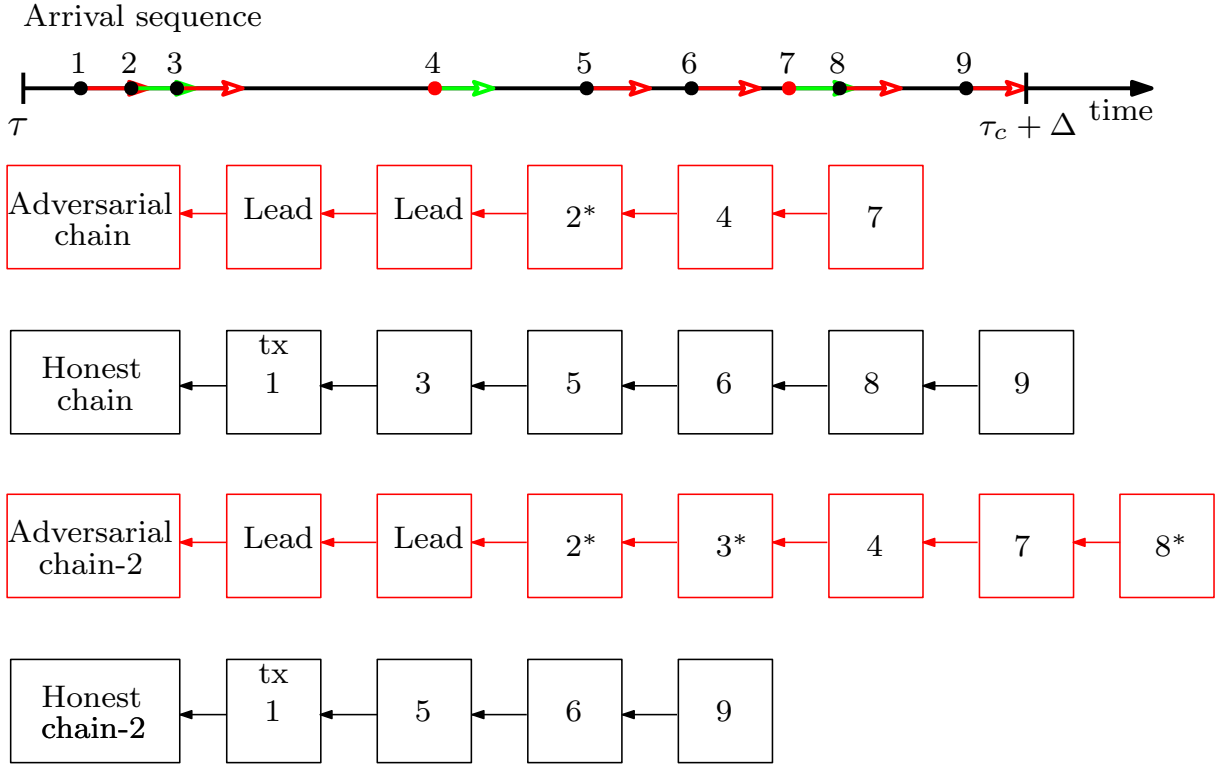}}
	\caption{A sample path of arrivals and rigged chains}
	\label{rigged-seqn}
\end{figure}

In \cite{guo-btc-sec-lat} every honest arrival that is a tailgater, i.e., arrivals within $\Delta$ of any other arrival, is converted to an adversarial block. However, this need not be true for AHBODH to hold. For example, an honest arrival need not be converted if all arrivals within the preceding $\Delta$ are already adversarial. More specifically, consider the honest chain to be consisting of only honest jumper blocks and all other blocks that are mined belong to the adversarial chain, thus, AHBODH holds, and the transition matrix during pre-mining is as follows,
\begin{align}
    \bar{P}=\begin{bmatrix}
        \bar{\alpha}_{0} & \bar{\alpha}_{1}+\beta & \bar{\alpha}_{2}          & \bar{\alpha}_{3} & \bar{\alpha}_{4} & \ldots\\
        \bar{\alpha}_{0} & \bar{\alpha}_{1}       & \bar{\alpha}_{2}+\beta    & \bar{\alpha}_{3} & \bar{\alpha}_{4} & \ldots\\
        0          & \bar{\alpha}_{0} & \bar{\alpha}_{1}       & \bar{\alpha}_{2}+\beta    & \bar{\alpha}_{3} & \ldots\\
        0          & 0 & \bar{\alpha}_{0} & \bar{\alpha}_{1}       & \bar{\alpha}_{2}+\beta    &  \ldots\\
        \vdots&\vdots&\vdots&\vdots&\vdots&\ddots\\
    \end{bmatrix}
\end{align}
Note the difference between $P$ and $\bar{P}$: When we upper bound the lead, we toss a coin, if it lands honest,  each arrival during $\Delta$ delay interval that follows this block is considered as adversarial, whereas, we discarded honest blocks arriving in the $\Delta$ delay interval when we lower bounded the lead. Hence, all blocks except jumpers belong to the adversarial chain. 

As an example, consider the arrival sequence shown in \figref{rigged-seqn} when confirmation interval begins with $\tau$ and ends with $\tau_c+\Delta$ (for the honest chain on top) where $k=6$. Black dots represent honest arrivals and red dots represent adversarial arrivals. Red arrows represent $\Delta$ delay intervals of jumper honest blocks and green arrows represent $\Delta$ delay intervals of any other type of block. If we only consider arrivals happening in red arrows to be rigged by the adversary, we get the adversarial chain and honest chain shown on top (block $2$ gets rigged). If we further consider arrivals in green regions to be rigged as well (like it was done in \cite{guo-btc-sec-lat}), then, we get the two chains on the bottom (block $3$ and $8$ get rigged in addition to block $2$). Clearly, adversary wins in the second scenario but not in the first, hence the potential improvement. 

The steady state of matrix $\bar{P}$ can be found by Ramaswami's formula \cite{ramaswami-mg1}, as $\bar{P}$ is a stochastic matrix of M/G/1 type.

\begin{lemma}\label{ramaswami-lemma}
     For $1>2\beta+\alpha\lambda\Delta$, steady state of $\Bar{P}$ can be found recursively using 
    \begin{align}
        \pi_i&=\frac{\pi_0 \Bar{b}_i+\sum_{j=1}^{i-1}\pi_j \Bar{a}_{i+1-j}}{1-\Bar{a}_1}, \quad i\geq1 \label{ramaswami-form}
    \end{align}
    where
    $\pi_0=\frac{1-2\beta-\alpha\lambda\Delta}{\alpha}$, $\Bar{a}_i=\sum_{j\geq i}\Bar{\alpha}_j + \beta\cdot\mathbbm{1}_{i\leq 2}$ and $\Bar{b}_i=\sum_{j\geq i}\Bar{\alpha}_j + \beta\cdot\mathbbm{1}_{i\leq 1}$.
\end{lemma} 
\begin{Proof}
    Note that, finding $\Bar{a}_i$ and $\Bar{b}_i$ is a straightforward application of Ramaswami's formula and they are provided in Table~\ref{freq-table}. Here, we only prove $\pi_0=\frac{1-2\beta-\alpha\lambda\Delta}{\alpha}$, which is done by summing each hand side of the Ramaswami equation for $i\geq1$ in order to express $\pi_i$ in terms of $\pi_0$.
    
    Summing \eqref{ramaswami-form} for $i\geq1$, we get
    \begin{align}
        (1-\Bar{a}_1)\sum_{i\geq1}\pi_i&=\sum_{i\geq1}\left(\pi_0 \Bar{b}_i+\sum_{j=1}^{i-1}\pi_j \Bar{a}_{i+1-j}\right)\\
        &=\pi_0\sum_{i\geq1}\Bar{b}_i+\sum_{i\geq2}\sum_{j=1}^{i-1}\pi_j \Bar{a}_{i+1-j}\\
        &=-\pi_0\beta+\pi_0\sum_{i\geq1}\Bar{a}_i+\sum_{j\geq1}\sum_{i>j}\pi_j \Bar{a}_{i+1-j}\\
        &=-\pi_0\beta+\pi_0\sum_{i\geq1}\Bar{a}_i+\sum_{j\geq1}\pi_j\sum_{i\geq2} \Bar{a}_{i}
    \end{align}
    Substituting $\sum_{i\geq1}\pi_i=1-\pi_0$ gives
    \begin{align}
        (1-\Bar{a}_1)(1-\pi_0)=-\pi_0\beta+\pi_0\sum_{i\geq1}\Bar{a}_i+(1-\pi_0)\sum_{i\geq2} \Bar{a}_{i}\\
    \end{align}
    and moving terms, we have
    \begin{align}
    \alpha\pi_0&=1-\sum_{i\geq1} \Bar{a}_{i}\\
    &=1- 2\beta-\sum_{i\geq1} \sum_{j\geq i}\Bar{\alpha}_j \\
    &=1- 2\beta-\sum_{j\geq1}\sum_{i=1}^{j}\Bar{\alpha}_j\\
    &=1- 2\beta-\sum_{j\geq1}j\Bar{\alpha}_j \label{lower-ramas-exp}\\
    &=1- 2\beta-\alpha\lambda\Delta
    \end{align}
    gives the desired result.
\end{Proof}

We denote the upper bound on the lead as $\bar{L}$ and the PMF of $\bar{L}$ with $P'_{1}$. Note that with the random walk presented in $\Bar{P}$, the random walk escapes to infinity when $1\leq 2\beta+\alpha\lambda\Delta$. Hence, the maximum provable fault tolerance for the rigged model using $\Bar{P}$ is $1>2\beta+\alpha\lambda\Delta$, whereas the ultimate fault tolerance is $1>2\beta+\alpha\beta\lambda\Delta$.

\subsection{Confirmation Interval}
 As the transaction enters the mempool at $\tau$, it will be included in the next honest block to be mined. However, if $\tau$ is in the $\Delta$ delay immediately following a jumper that arrived before $\tau$, the honest blocks mined during this $\Delta$ delay are rigged. Hence, since the $tx$ will be included in the first jumper block after $\tau$, without loss of generality, we assume again that $tx$ enters the mempool at a time $\tau$, where the first honest block to be mined contains the $tx$ and is a jumper. Further, this implies that the steady-state distribution of $\Bar{P}$ is an upper bound for the adversarial lead. Similar to the case of the lower bound we find the number of blocks mined before each honest jumper is published. Keep in mind that, in addition to delaying the publication of jumpers by $\Delta$, all honest blocks mined during this delay are rigged. Hence, all blocks but jumpers are adversarial in this model.

\begin{lemma}\label{lemma-upper-geo}
Under the $\Delta$ delay model, the number of blocks mined between the publication times of two consecutive jumper blocks (except the jumper) $\bar{C}$ has the following distribution,
\begin{align}
    P_{\bar{C}}(c)=\alpha\beta^{c}e^{-\lambda\Delta}\sum_{j=0}^{c}\frac{(\lambda\Delta/\beta)^j}{j!}
    \label{upper_geom_eq}
\end{align}
\end{lemma}
\begin{Proof}
After an honest (jumper) is published, we know that the probability of having $m$ adversarial arrivals before the next honest arrival (which is a jumper by definition) is $\beta^m\alpha$. However, after this honest block arrives, there is a $\Delta$ delay interval until it is published, during which all honest arrivals will be converted to adversarial, hence there can be additional $j$ adversarial arrivals w.p.~$e^{-\lambda\Delta}\frac{(\lambda\Delta)^j}{j!}$. Combining these two results, i.e., $c=j+m$, we obtain \eqref{upper_geom_eq}. Clearly, at the end of the $\Delta$ delay interval, this jumper will be published, and thus, the next jumper also has the same distribution and they are independent by the memoryless property, i.e., they are i.i.d.
\end{Proof}
\begin{corollary}\label{corollary-upper-pascal}
The number of blocks mined during the confirmation interval (except jumpers) under the rigged model $\bar{S}$ has the following distribution 
\begin{align}
    P_{\bar{S}}(s) &= \bar{\alpha}_{0}^{k}\sum_{n=0}^{s}\binom{k\!-\!1\!+\!n}{n}\frac{(\lambda\Delta k)^{s-n}}{(s-n)!}\beta^n \label{upper-pascal-like-eqn}
\end{align}
\end{corollary}

\begin{Proof}
All jumpers including the first has the distribution of $\bar{C}$ in (\ref{upper_geom_eq}). Further, $\bar{S}$ is the sum of $k$ i.i.d.~random variables each distributed as $\bar{C}$. First, we find the probability generating function (PGF) of $\bar{C}$, i.e., $G_{\bar{C}} (z)=\mathbb{E}[z^{\bar{C}}]$, as
\begin{align}
    G_{\bar{C}}(z)&=\alpha e^{-\lambda\Delta}\sum_{i\geq0}(z\beta)^{i}\sum_{j=0}^{i}\frac{(\lambda\Delta/\beta)^j}{j!}\\
    &=\bar{\alpha}_{0}\sum_{j\geq0}\sum_{i\geq j}(z\beta)^{i}\frac{(\lambda\Delta/\beta)^j}{j!}\\
    &=\bar{\alpha}_{0}\sum_{j\geq0}\frac{(z\lambda\Delta)^j}{(1-z\beta)\cdot j!}\\
    &=\bar{\alpha}_{0}\frac{e^{z\lambda\Delta}}{1-z\beta}
\end{align}
Next, we find the PGF of $\bar{S}$, as
\begin{align}
    G_{ \bar{S}}(z) &= G_{\bar{C}}(z)^{k}\\
    &=\bar{\alpha}_{0}^{k}\left(\frac{e^{z\lambda\Delta}}{1-z\beta}\right)^{k} \label{u-intermed1}\\
    &=\bar{\alpha}_{0}^{k}\sum_{n\geq0}\binom{k-1\!+\!n}{n}(z\beta)^{n}\sum_{m\geq0}\frac{(z\lambda\Delta k)^m}{m!} \label{u-intermed2}\\
    &=\bar{\alpha}_{0}^{k}\sum_{m\geq0}\sum_{n\geq0}\binom{k\!-\!1\!+\!n}{n}\frac{(\lambda\Delta k)^m}{m!}z^{n+m}\beta^n\\
    &=\sum_{s\geq0}\bar{\alpha}_{0}^{k}z^{s}\sum_{n=0}^{s}\binom{k\!-\!1\!+\!n}{n}\frac{(\lambda\Delta k)^{s-n}}{(s-n)!}\beta^n \label{u-intermed3}
\end{align}
where in going from (\ref{u-intermed1}) to (\ref{u-intermed2}), we use the identity in (\ref{pascal}).
The desired result follows from (\ref{u-intermed3}). 
\end{Proof}

We denote the PMF of $\bar{S}$ with $P'_{2}$.

\subsection{Post-Confirmation Race}
Clearly, if $\bar{L}+\bar{S}\geq k$, then, the adversary wins the race, otherwise, the deficit of the adversary is $\bar{D}=k-\bar{L}-\bar{S}$, which it has to make up during the post-confirmation race. 

Here, we consider two arrivals at a time which brings improvement to the random walk considered in \cite{guo-btc-sec-lat}. In the rigged model of \cite{guo-btc-sec-lat}, an arrival is i.i.d.~Bernoulli, and coin toss lands honest w.p.~$\alpha e^{-\lambda\Delta}$. Let $(H,A)$ denote the growth of honest and adversarial chains after two arrivals. If we consider two consecutive i.i.d.~coin tosses separately and sum their results, then,
\begin{enumerate*}
    \item $\bar{E}_1'$ denoting $(2,0)$ happens w.p.~$(\alpha e^{-\lambda\Delta})^2$;
    \item $\bar{E}_2'$ denoting $(1,1)$ happens w.p.~$2(\alpha e^{-\lambda\Delta})(1-\alpha e^{-\lambda\Delta})$; and
    \item $\bar{E}_3'$ denoting $(0,2)$ happens w.p.~$(1-\alpha e^{-\lambda\Delta})^2$.
\end{enumerate*}
In our analysis $\bar{E}_1'$ stays the same but we decrease the probability of $\bar{E}_3'$ by considering two consecutive coin tosses together instead of considering them separately.

Without loss of generality, assume that we start from time zero and consider groups of two arrivals in the rigged model. Note that the genesis block is the zeroth jumper that arrives at time zero. Here, $\bar{E}_1$, i.e., $(H,A)=(2,0)$ has the same probability as $\bar{E}_1'$. For $\bar{E}_2$ to happen, i.e., $(H,A)=(1,1)$, there are $3$ cases:

\begin{enumerate}
\item The first arrival, arriving at $t_1>\Delta$, is honest and the second arrival, arriving at $t_2$ is not an honest block or $t_2-t_1<\Delta$ (gets rigged if honest). This case is treated in the same way as it is done in \cite{guo-btc-sec-lat} and has probability $\alpha e^{-\lambda\Delta}(1-\alpha e^{-\lambda\Delta})$. \label{E_2-item-11}
\item The first arrival, arriving at $t_1>\Delta$, is adversary and the second arrival is honest. This case, has probability $\alpha\beta e^{-\lambda\Delta}$. \label{E_2-item-12}
\item The first arrival arrives at $t_1<\Delta$ (gets rigged if honest), the second arrival is honest and $t_2>\Delta$. This case has probability $\alpha\lambda\Delta e^{-\lambda\Delta}$. \label{E_2-item-13}
\end{enumerate}

We note that, in all these cases, the honest block has to be on different height than all previous (non-rigged) honest blocks hence, satisfying AHBODH. Moreover, this improves the results of \cite{guo-btc-sec-lat} in certain scenarios. For example, if the first arrival with $t_1>\Delta$ is adversary and the second arrival is honest with $t_2-t_1<\Delta$, then, \cite{guo-btc-sec-lat} converts the second arrival to adversary, we do not.

After the second arrival at $t_2$, by the memorylessness property of exponential arrivals, same arguments will hold for the next group of  two arrivals and so forth. Hence, each group is identically distributed and independent from each other. Note that,  $\mathbb{P}(\bar{E}_2)=\rho\geq \mathbb{P}(\bar{E}_2')$, hence the improvement.

It is straightforward to show that a random walk which starts at the origin and moves two steps at a time, left or right with $\mathbb{P}(\bar{E_1})=\bar{\alpha}^2$ and $\mathbb{P}(\bar{E_3})=\bar{\beta}^2$, respectively, can reach the point $2m$ with probability
\begin{align}
    \mathbb{P}(\bar{M}\geq2m)=\left(\frac{\bar{\beta}}{\bar{\alpha}}\right)^{2m} \label{2-step-geo}
\end{align}
where $\bar{M}$ denotes the maximum point this random walk ever reaches. Let $\bar{M}'$ denote the maximum difference between the heights of adversarial and honest chains under the simplified $(A,H)$ random walk model with two arrivals at a time. Thus, if $\bar{D}$ is even, then we have $\mathbb{P}(\bar{M}\geq \bar{D})=\mathbb{P}(\bar{M}'\geq \bar{D})$. If odd, however, there can be cases of unobservable adversarial win at the end of two tosses. To avoid this complication, we simply consider a single toss (using the rigged model of \cite{guo-btc-sec-lat}) initially to make sure deficit becomes even before using \eqref{2-step-geo}: $\mathbb{P}(\bar{M}'\geq \bar{D})=\bar{\alpha}_{0}\mathbb{P}(\bar{M}\geq \bar{D}+1)+(1-\bar{\alpha}_{0})\mathbb{P}(\bar{M}\geq \bar{D}-1)$. We denote CDF of $\bar{M}'$ with $F'_{3}$.

\begin{theorem} \label{upper-theorem}
Given mining rate $\lambda$, honest fraction $\alpha$, delay bound $\Delta$ and confirmation depth $k$, a confirmed transaction cannot be discarded w.p.~greater than:
\begin{align}
    1-\sum_{i+j<k} P'_1(i)P'_2(j)F'_3(k-1-i-j) \label{thm-2 eq}
\end{align}
where 
\begin{align*}
    P'_1(0)=\frac{1-2\beta-\alpha\lambda\Delta}{\alpha},
\end{align*}     
\begin{align*}
    P'_1(i)=\frac{P'_1(0) \Bar{b}_i+\sum_{j=1}^{i-1}P'_1(j) \Bar{a}_{i+1-j}}{1-\Bar{a}_1}, \quad i\geq1,
\end{align*}
\begin{align*}
    P'_{2}(j) = \bar{\alpha}_{0}^{k}\sum_{n=0}^{j}\binom{k\!-\!1\!+\!n}{n}\frac{(\lambda\Delta k)^{j-n}}{(j-n)!}\beta^n,\quad j\geq 0,
\end{align*} 
\begin{align*}
    F'_{3}(l)=1-\left(\frac{\bar{\beta}}{\bar{\alpha}}\right)^{l+1}, \quad \text{(l odd)},
\end{align*}
\begin{align*}
    F'_{3}(l)=1-\left(\frac{\bar{\beta}}{\bar{\alpha}}\right)^{l}\left(1-\bar{\alpha}+\bar{\alpha}\left(\frac{\bar{\beta}}{\bar{\alpha}}\right)^{2}\right), \quad \text{(l even)}.\\
\end{align*}
where $\bar{\beta}<\bar{\alpha}$.
\end{theorem}

Although the maximum provable fault tolerance for the rigged model using $\Bar{P}$ is $1>2\beta+\alpha\lambda\Delta$, as we use a simpler model in the post-confirmation race when we find the upper bound where adversary has even more advantage, $\bar{\beta}<\bar{\alpha}$ is the fault tolerance of Theorem~\ref{upper-theorem}.

\section{Numerical Results}
\begin{figure}[t]
	\centerline{\includegraphics[width=0.6\columnwidth]{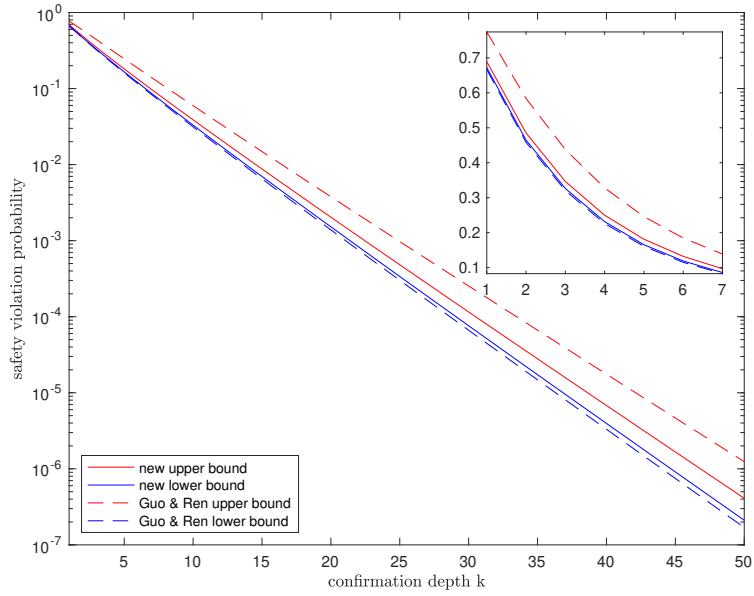}}
	\caption{Bitcoin safety violation with $\alpha=0.75$.}
	\label{btc-0.75-fig}
\end{figure}

\begin{figure}[ht!]
	\centerline{\includegraphics[width=0.6\columnwidth]{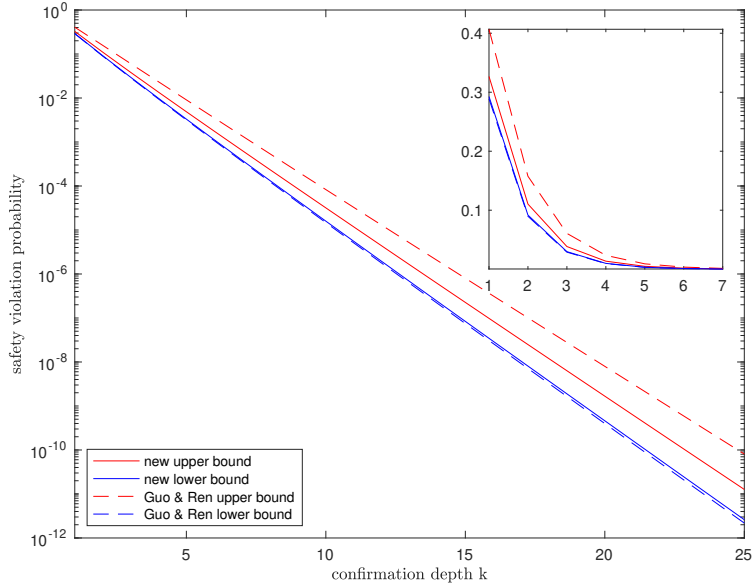}}
	\caption{Bitcoin safety violation with $\alpha=0.90$.}
	\label{btc-0.90-fig}
\end{figure}

We present our results for Bitcoin in \figref{btc-0.75-fig} and \figref{btc-0.90-fig}. We choose $\lambda=1/600$ and $\Delta=10$ seconds \cite{DSN-Bitcoin-Monitoring} for $\alpha=0.75$ and $\alpha=0.90$, respectively. We compare the tightest result of \cite{guo-btc-sec-lat} (Theorem~$3$) with our tightest result. Note that, lower and upper bounds in our results have the same order of magnitude. As $k$ grows, i.e., the confirmation time becomes longer, and non-jumper honest arrivals grow, which in turn get rigged and increase the adversarial win probability. Hence, the difference in the slopes of upper and lower bounds. We further note that, the improvements we introduce for pre-mining and post-confirmation regions mostly shift the curve, whereas the improvement we introduce in the confirmation interval changes the slopes. These results narrow down the safety violation probability of Bitcoin under $6$-block confirmation rule with $\alpha=0.75$ to $[0.12,0.13]$, and with $\alpha=0.90$ to $[0.00112, 0.00173]$ (as mentioned in Section~\ref{intro-sec}).


\begin{figure}[t]
    \centerline{\includegraphics[width=0.6\columnwidth]{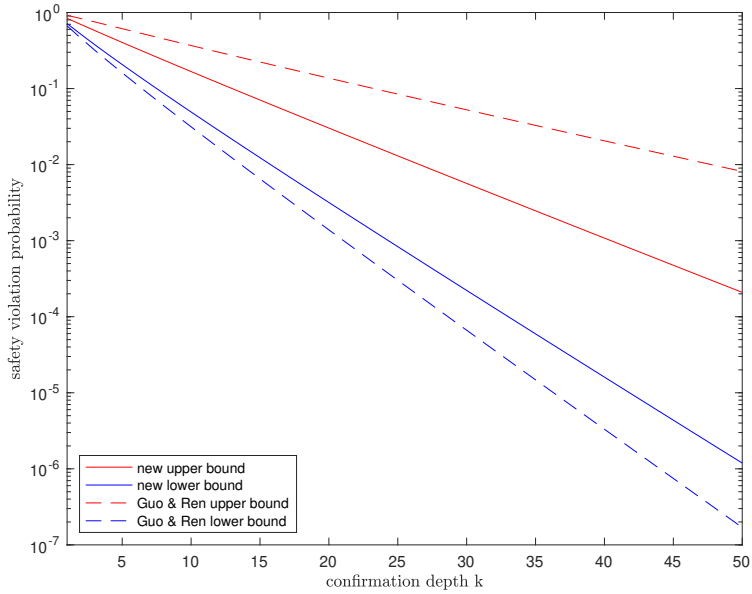}}
	\caption{Ethereum safety violation with $\alpha=0.75$.}
	\label{eth-75-2-fig}
\end{figure}

We present our results for Ethereum in \figref{eth-75-2-fig}. We choose $\lambda=1/13$ and $\Delta=2$ seconds and $\alpha=0.75$. Note how increasing $\lambda\Delta$ affects the improvement. As also observed in \cite{guo-btc-sec-lat}, the safety violation probability is mainly determined by the confirmation interval, hence, modeling the honest chain with jumpers which heavily depends on the value of $\lambda\Delta$, we are able improve the bounds by orders of magnitude.

\begin{figure}[ht!]
	\centerline{\includegraphics[width=0.6\columnwidth]{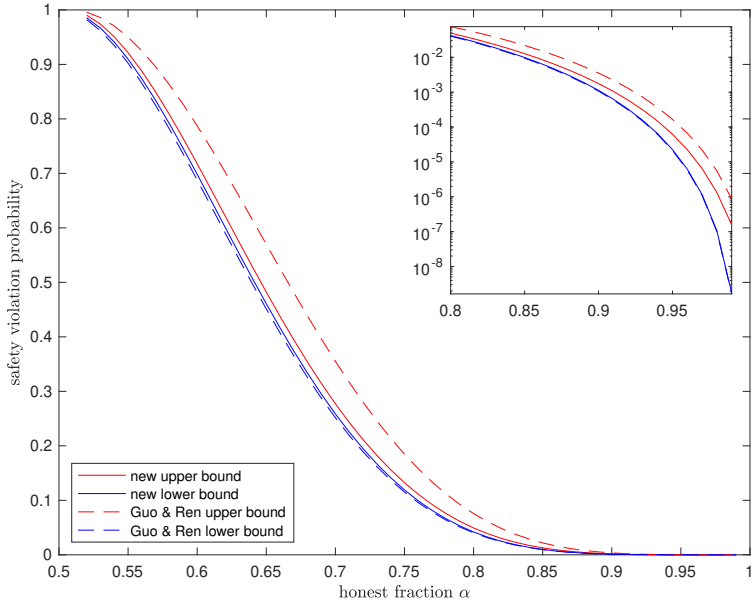}}
	\caption{Bitcoin safety violation vs $\alpha$ under 6-block confirmation rule.}
	\label{btc_perc_vs_sec}
\end{figure}

We present Bitcoin's security bounds for varying honest fraction $\alpha\in[0.52,0.99]$ under $6$-block confirmation rule in \figref{btc_perc_vs_sec}. Taking all numerical results into account, we see the following trends: As $\alpha$ grows, the difference between the safety violation probability upper and lower bounds grows in orders of magnitude. We were able to decrease this effect by our improvements on the the upper bound, whereas for small $\Delta$, our improvements on the lower bound are modest. However, as our improvements on the lower bound focus on $\Delta$ delay strategy, as $\lambda\Delta$ grows, e.g., Ethereum, then the improvements on the lower bound become more significant.  

\section{Conclusion}
In this paper, we improved state-of-the-art security-latency bounds for blockchains  by considering the $\Delta$ delay strategy and tightening the rigged model of \cite{guo-btc-sec-lat}. In order to analyze the $\Delta$ delay strategy, we introduced a new Markov chain and a random walk that model the statistical behavior of the race between the adversarial and honest chains. As the confirmation interval is the most crucial part of this race, we derived the distribution of the number of adversarial blocks generated during the confirmation interval under the new model, which lead to considerable improvements on the lower bound provided in \cite{guo-btc-sec-lat} under large $\Delta$. We also showed that, with some adjustments the same random walk  represents the rigged model during the confirmation interval, which lead to even stronger results on the upper bound. For the pre-mining gain and the post-confirmation race, we simplified the Markov chains to analytically more tractable ones, which further improved the bounds of \cite{guo-btc-sec-lat} by shifting the curves (in logarithmic representation).

We remark that our bounds are in open form. Closed form expressions can give a better understanding of the dependencies between parameters. We leave this further research for future work. For the $\Delta$ delay strategy and rigged model that we considered in this paper, we believe that our improvements in the confirmation interval is quite tight. However, one may come up with adversarial strategies that further tighten the bounds. Also there is some room for further improvements in post-confirmation races under the $\Delta$ delay strategy and the rigged model.

\appendix
\section{Appendix}\label{app-a}
\begin{lemma} 
    {\normalfont \textbf{(Ramaswami\cite{ramaswami-mg1})}} For $1\geq \beta(2+\alpha\lambda\Delta)$, steady state of $P$ can be found recursively using 
    \begin{align}
        \pi_i&=\frac{\pi_0 b_i+\sum_{j=1}^{i-1}\pi_j a_{i+1-j}}{1-a_1}, \quad i\geq1 
    \end{align}
    where
    $\pi_0=\frac{1-\beta(2+\alpha\lambda\Delta)}{\alpha}$.
\end{lemma}

\begin{Proof}
    The proof follows the same steps as in Lemma~\ref{ramaswami-lemma}, where we replace $\Bar{a}_i$, $\Bar{b}_i$ and $\Bar{\alpha}_i$ with $a_i$, $b_i$ and $\alpha_i$. After \eqref{lower-ramas-exp}, we end up with $\alpha\beta\lambda\Delta$ instead of $\alpha\lambda\Delta$, since we consider just adversarial arrivals for each $\Delta$ delay interval instead of all arrivals.
\end{Proof}

The resulting lead from $P'$ is a lower bound on the lead obtained from the steady state of $P$ as we truncate the number of adversarial arrivals during a $\Delta$ delay interval to at most two arrivals. Note that, here, when we say the lead distribution $L'$ lower bounds the lead distribution $L$, we mean $P(L<i)\leq P(L'<i)$, for $\forall i$. 

Note that, with the random walk presented in $P$, the random walk escapes to infinity when $1\leq \beta(2+\alpha\lambda\Delta)$ which is the same condition as $\beta\geq\frac{1-\beta}{1+(1-\beta)\lambda\Delta}$, i.e., the converse of the ultimate fault tolerance condition. This is not surprising, since $P$ presents the $\Delta$ delay strategy, where the honest chain grows at the slowest possible rate, i.e., $\frac{1}{\Delta+1/\alpha\lambda}$, and is in a race with adversarial chain that has the rate $\beta\lambda$.

In our figures, we use the steady states of $P$ and $\bar{P}$ consistently. We note that the bounds of the two lower bounds visually overlap with each other, hence we refrain from comparing them with plots. As an example, for Bitcoin $6$-block confirmation rule with $\alpha=0.9$, using $P'$ results in safety violation lower bound probability of $0.00112412$ whereas $P$ gives $0.00112415$.

\bibliographystyle{ieeetr}
\bibliography{lib}

\begin{thebibliography}{10}

\bibitem{btc-whitepaper}
S.~Nakamoto, ``Bitcoin: A peer-to-peer electronic cash system.''
  https://bitcoin.org/bitcoin.pdf, March 2008.

\bibitem{garay}
J.~Garay, A.~Kiayias, and N.~Leonardos, ``The bitcoin backbone protocol:
  Analysis and applications,'' in {\em EUROCRYPT 2015}, April 2015.

\bibitem{non-lockstep-sec-lat}
R.~Pass and E.~Shi, ``Rethinking large-scale consensus,'' in {\em IEEE CSF
  2017}, August 2017.

\bibitem{guo-close-sec-lat}
J.~Li, D.~Guo, and L.~Ren, ``Close latency-security trade-off for the nakamoto
  consensus,'' in {\em ACM AFT 2021}, November 2021.

\bibitem{guo-btc-sec-lat}
D.~Guo and L.~Ren, ``Bitcoin's latency–security analysis made simple,'' in
  {\em ACM AFT 2022}, September 2023.

\bibitem{gazi-sec-lat}
P.~Gazi, L.~Ren, and A.~Russell, ``Practical settlement bounds for
  proof-of-work blockchains,'' in {\em ACM CCS 2022}, November 2022.

\bibitem{nakamoto-always-wins}
A.~Dembo, S.~Kannan, E.~Tas, D.~Tse, P.~Viswanath, X.~Wang, and O.~Zeitouni,
  ``Everything is a race and nakamoto always wins,'' in {\em ACM CCS 2020},
  November 2020.

\bibitem{ramaswami-mg1}
V.~Ramaswami, ``A stable recursion for the steady state vector in markov chains
  of m/g/1 type,'' {\em Communications in Statistics. Stochastic Models},
  vol.~4, no.~1, pp.~183--188, 1988.

\bibitem{ross-stochastic}
S.~Ross, {\em Stochastic processes}.
\newblock John Wiley \& Sons, Inc., second~ed., 1996.

\bibitem{DSN-Bitcoin-Monitoring}
``{DSN Bitcoin Monitoring}.'' \url{https://dsn.tm.kit.edu/bitcoin/}.
\newblock Accessed: 2021-07-30.

\end{thebibliography}

\end{document}